# Development of virtual morphometric globes using Blender

I.V. Florinsky*, S.V. Filippov

Institute of Mathematical Problems of Biology, Russian Academy of Sciences
Pushchino, Moscow Region, 142290, Russia

**Abstract**

Virtual globes – programs implementing interactive three-dimensional (3D) models of planets – are increasingly used in geosciences. Global morphometric models can be useful for tectonic and planetary studies. We describe the development of the first testing version of the system of virtual morphometric globes for the Earth, Mars, and the Moon. As the initial data, we used three 15'-gridded global digital elevation models (DEMs) extracted from SRTM30_PLUS, the Mars Orbiter Laser Altimeter, and the Lunar Orbiter Laser Altimeter gridded archives. For three planetary bodies, we derived global digital models and maps of several morphometric attributes (i.e., horizontal curvature, vertical curvature, minimal curvature, maximal curvature, and catchment area). To develop the system, we used Blender, the open-source software for 3D modeling and visualization. First, a 3D sphere model was generated. Second, the global morphometric maps as textures were imposed to the sphere surface. Finally, the real-time 3D graphics Blender engine was used to implement globe rotation and zooming. The testing of the developed system demonstrated its good performance. Morphometric globes clearly represent peculiarities of planetary topography, according to the physical and mathematical sense of a particular morphometric variable.

## 1. Introduction

In the last years, significant progress has been made in the development and application of virtual globes (Tuttle et al., 2008; Blaschke et al., 2012). Virtual globe are programs implementing interactive three-dimensional (3D) models of planetary bodies. The list of virtual globes includes World Wind (NASA, 2003–2011), Earth3D (Gunia, 2004–2015), Google Earth (Google Inc., 2005–2014), Marble (KDE, 2007–2014), Cesium (AGI, 2012–2015), and others. Virtual globes enable to carry out 3D multi-scale visualization of complex spatially distributed multi-layer data with capabilities to move around the globe and to change the user viewing angle and position relative to the globe. Virtual globes are increasingly used to solve various multi-scale tasks of geosciences (Ryakhovsky et al., 2003; Chen and Bailey, 2011; Paraskevas, 2011; Guth, 2012; Yu and Gong, 2012; Zhu et al., 2014).

Topography is one of the main factors controlling processes taking place in the near-surface layer of the planet. In particular, topography is one of the soil forming factors. At the same time, being a result of the interaction of endogenous and exogenous processes of different scales, topography can reflect the geological structure of a terrain. In this connection, digital terrain models (DTMs) – two-dimensional discrete functions of morphometric variables – are widely used to solve various multiscale problems of geomorphology, hydrology, remote sensing, soil science, geology, geophysics, geobotany, glaciology, oceanology, climatology, planetology, and other disciplines (Wilson and Gallant, 2000; Li et al., 2005; Hengl and Reuter, 2009; Florinsky, 2012). The list of morphometric variables (topographic attributes) includes: (a) local variables: horizontal curvature ($k_h$), vertical curvature ($k_v$), minimal curvature ($k_{min}$), maximal curvature ($k_{max}$), etc.; and (b) nonlocal variables: catchment area (*CA*), dispersive area (*DA*), etc. (definitions and physical interpretations of the morphometric attributes can be found elsewhere – Florinsky, 2012, Ch. 2).

---

* Correspondence to: iflorinsky@yahoo.ca

It is a common practice to visualize topography using hypsometric tinting and hill shading in virtual globes (Cozzi and Ring, 2011, Pt. 4) (Fig. 1). At the same time, specialized virtual morphometric globes still do not exist, although global morphometric models of the Earth and other planetary bodies can be useful for tectonic and planetary studies (Florinsky, 2008a, 2008b).

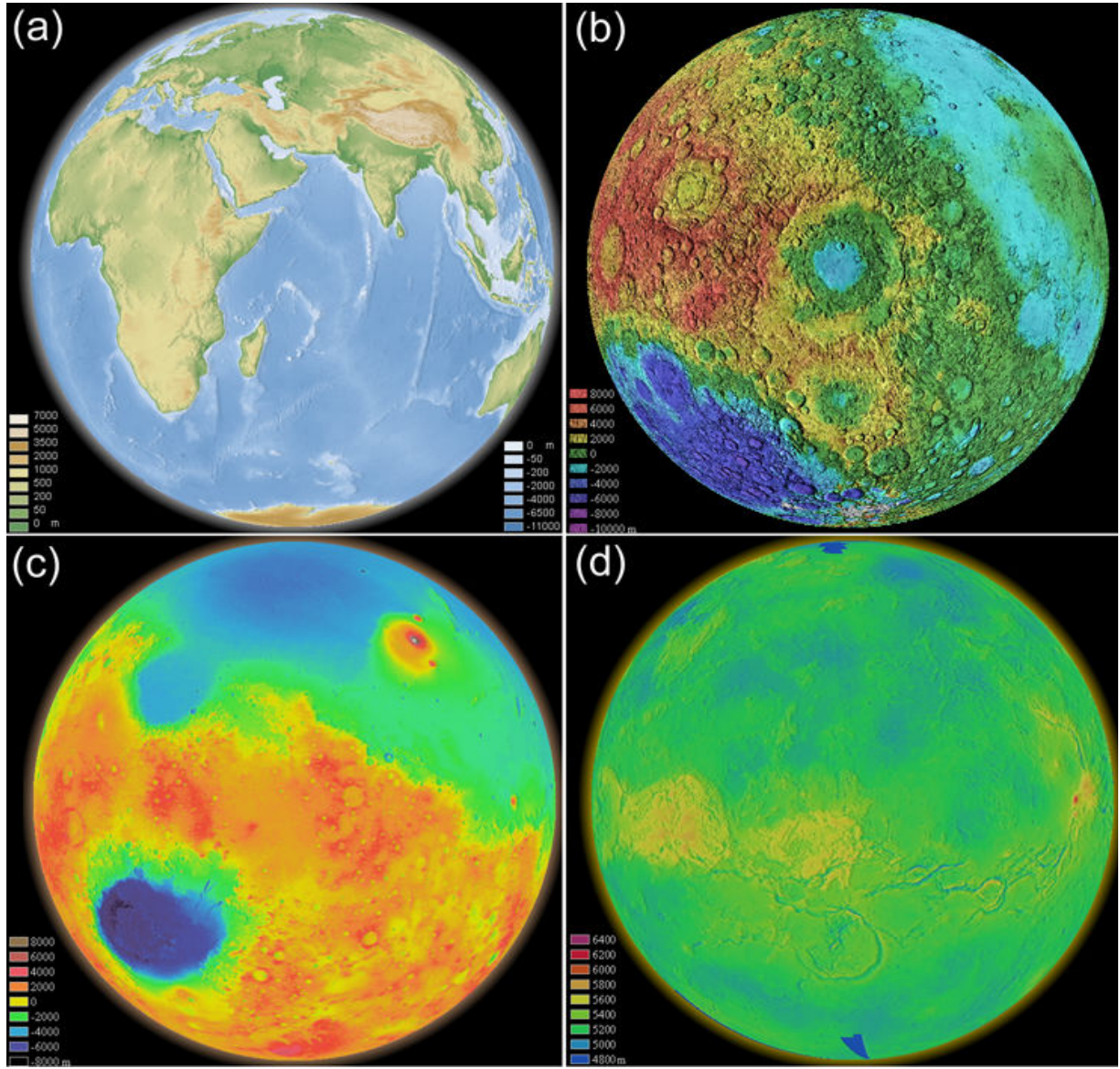

Fig. 1. Visualization of topography using hypsometric tinting and hill shading in Marble Virtual Globe 1.9.1 (KDE, 2007–2014): (a) the Earth; (b) the Moon, elevation data are from the Clementine gravity and topography data archive (Zuber et al., 1996) overlaid on the USGS shaded relief map; (c) Mars, elevation data are from the Mars Global Surveyor Laser Altimeter Mission Experiment gridded data record archive (Smith et al., 2003); and (d) Venus, elevation data are from the Magellan global topography, emissivity, reflectivity, and slope data record archive (Ford et al., 1992). Elevation maps are courtesy of the USGS Astrogeology Science Center.

Virtual globes utilize engines usually developed using 3D graphics application programming interface (API), such as OpenGL and WebGL (Cozzi and Ring, 2011). Rapid zooming of massive datasets is commonly provided by hierarchical tessellation of the globe surface (Mahdavi-Amiri et al., 2015). However, a level of complexity of rendering data should be considered in developing new virtual globes, in particular, in selection of the existing engine or development of a new one. Specialized multifunctional engines are not necessarily required for some relatively simple tasks.

Indeed, 3D scientific visualization can be carried out using the capabilities of the existing 3D graphics packages (Hansen and Johnson, 2005; Lipşa et al., 2012; Johnson and Hertig, 2014). In particular, Blender – the free and open-source software for 3D modeling and visualization (Blender Foundation, 2003–2015; Hess, 2010; Blain, 2012) – is currently applied for 3D scientific visualization (Kent, 2015), e.g., in biology (SciVis, 2011–2015; Autin et al., 2012), astronomy (Kent, 2013), and geoinformatics (Scianna, 2013).

In this paper we describe the development of the first testing version of the system of virtual morphometric globes for the Earth, Mars, and the Moon using Blender.

## 2. Data and methods

The development of the system consisted of two main steps:
1. Derivation of a set of global low-resolution morphometric maps of the Earth, Mars, and the Moon.
2. Generation of a 3D sphere model of a globe and integration of the morphometric maps with the 3D model.

### *2.1. Digital terrain modeling*

To facilitate the development of the first testing version of the system of virtual morphometric globes, we decided to work with low resolution DTMs. We used the following three 15'-gridded global digital elevation models (DEMs) as the initial data:
- A DEM of the Earth extracted from the global DEM SRTM30_PLUS (Sandwell et al., 2008; Becker et al., 2009).
- A DEM of Mars extracted from the Mars Orbiter Laser Altimeter (MOLA) gridded data record archive (Smith et al., 1999, 2003).
- A DEM of the Moon extracted from the Lunar Orbiter Laser Altimeter (LOLA) gridded data record archive (Neumann, 2008; Smith et al., 2010).

To suppress high-frequency noise, the DEMs were smoothed using the $3 \times 3$ moving window. The DEM of Mars was twice smoothed, and the DEMs of the Earth and the Moon were thrice smoothed.

For all three planetary bodies, we derived digital models of local morphometric attributes from the smoothed DEMs by the method for spheroidal equal angular grids (Florinsky, 1998; Florinsky, 2012, pp. 55–57). Digital models of nonlocal morphometric variables were calculated by the Martz – de Jong method (Martz and de Jong, 1988) adapted to spheroidal equal angular grids (Florinsky, 2012, pp. 60–61). To estimate linear sizes of spheroidal trapezoidal windows in DTM calculation and smoothing (Florinsky, 2012, pp. 57–58), standard values of the major and minor semiaxes of the Krasovsky ellipsoid and the Martian ellipsoid were used for the Earth and Mars, correspondingly; the Moon was considered as a sphere. The global DEMs were processed as virtually closed spheroidal matrices of elevations. Each global DTM included 1,036,800 points (the matrix $1440 \times 720$); the grid spacing was 15'.

Then we derived global maps of all calculated topographic attributes (Fig. 2). To deal with the large dynamic range of morphometric variables, we logarithmically transformed their digital models (Florinsky, 2012, p. 134). Data processing was performed by the software

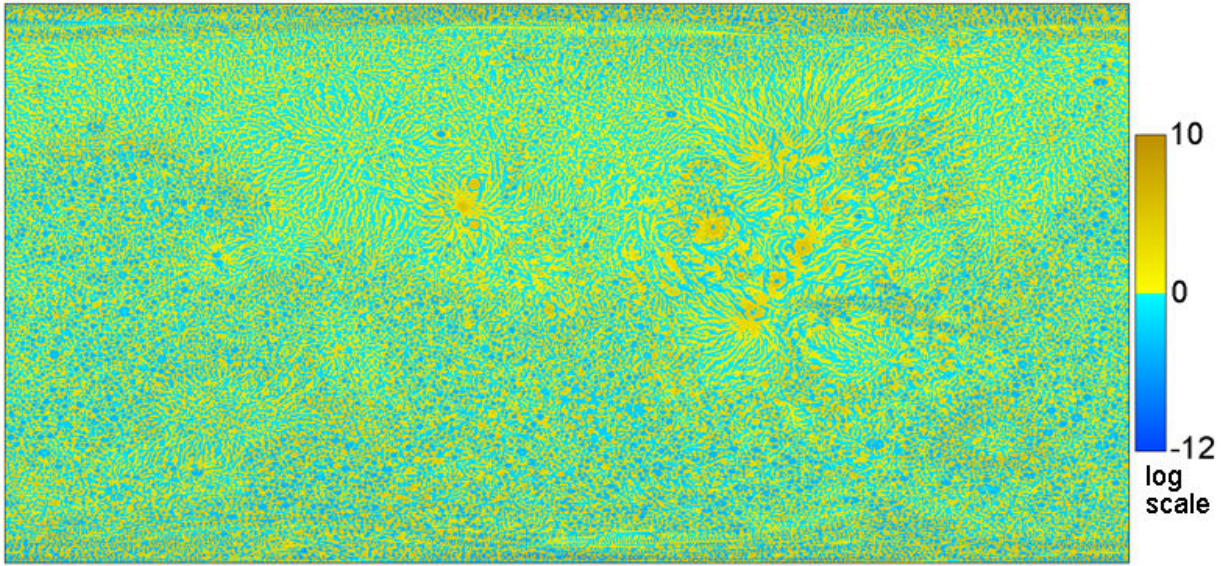

Fig. 2. An example of global morphometric maps: horizontal curvature for Mars.

LandLord (Florinsky, 2012, pp. 315–316). The global morphometric maps were saved as TIFF images for their subsequent use as textures in Blender.

## *2.2. Three-dimensional modeling*

To develop a testing version of a system of morphometric globes, we chose the package Blender 2.76b (Blender Foundation, 2015). It includes a real-time 3D graphics Blender Game Engine (BGE).

To construct the 3D model of the globe, we selected a UV sphere divided into 1152 rectangular polygons, that is, the 3D sphere is tessellated into 48 × 24 spherical trapezoids with sizes 7.5° × 7.5° (Fig. 3). A reasonably smooth representation of such a sphere can be achieved using the Phong shading model (Phong, 1975).

To create UV textures of the global morphometric maps, we processed TIFF images of the maps in the UV Editor, a part of the Blender package. Then, each UV morphometric texture was individually imposed to the sphere surface.

To create a latitude/longitude grid on the globe, we used the UV map of such a grid with the step of 7.5°. It was imposed as the second texture to the sphere surface, over a morphometric texture.

The globe rotation was performed using the mouse actuator embedded in BGE. To provide zoom, the BGE camera was connected with the logical chain of two sensors, mouse wheel down and mouse wheel up, which were linked with the motion actuators moving the BGE camera along the Y-axis (Fig. 4).

All morphometric globes were generated as individual scenes of Blender, which were then assembled into a single program.

## 3. Results and discussion

The real-time testing of the developed system demonstrated its good performance. Figures 5, 6, and 7 show examples (screenshots) of the developed morphometric globes for the Earth, Mars, and the Moon, correspondingly.

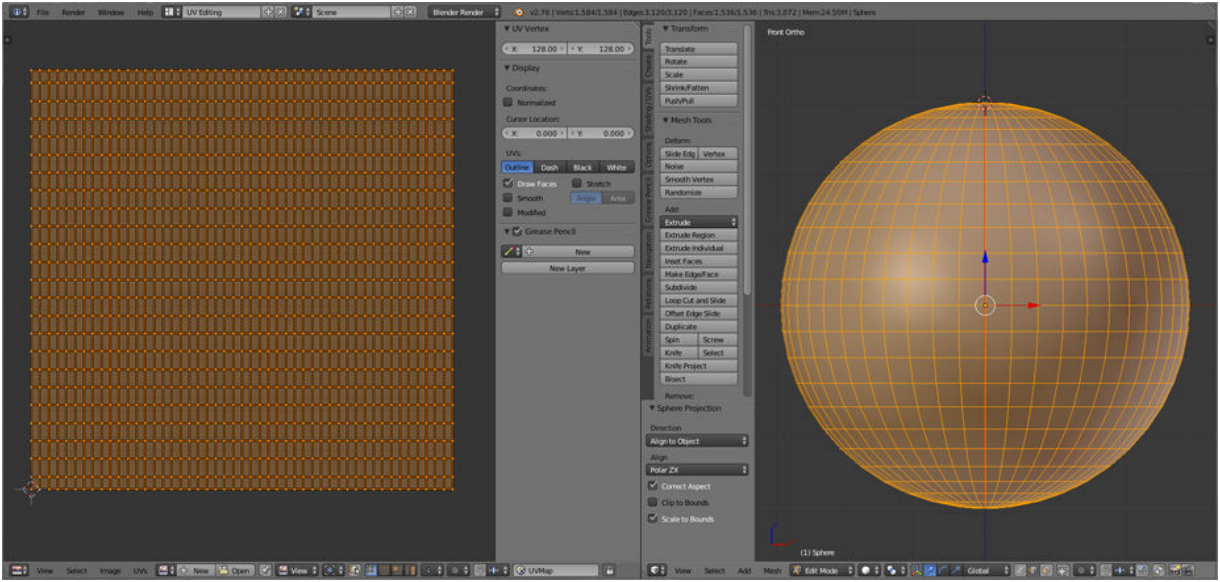

Fig. 3. A geometric model for a virtual globe: 3D model of a sphere (right) and UV-map of a sphere (left).

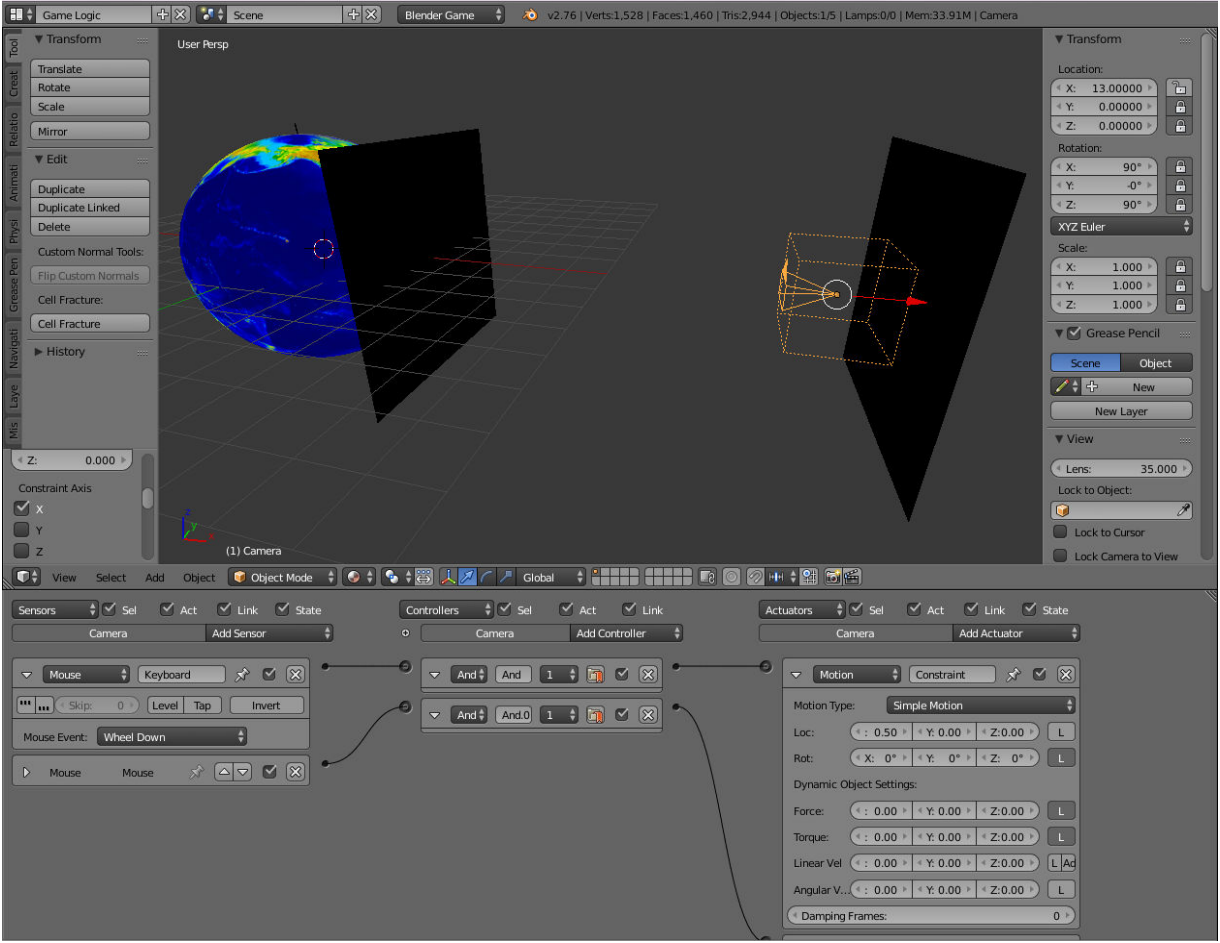

Fig. 4. The general view of a scene (upper) and the logic of zooming (lower).

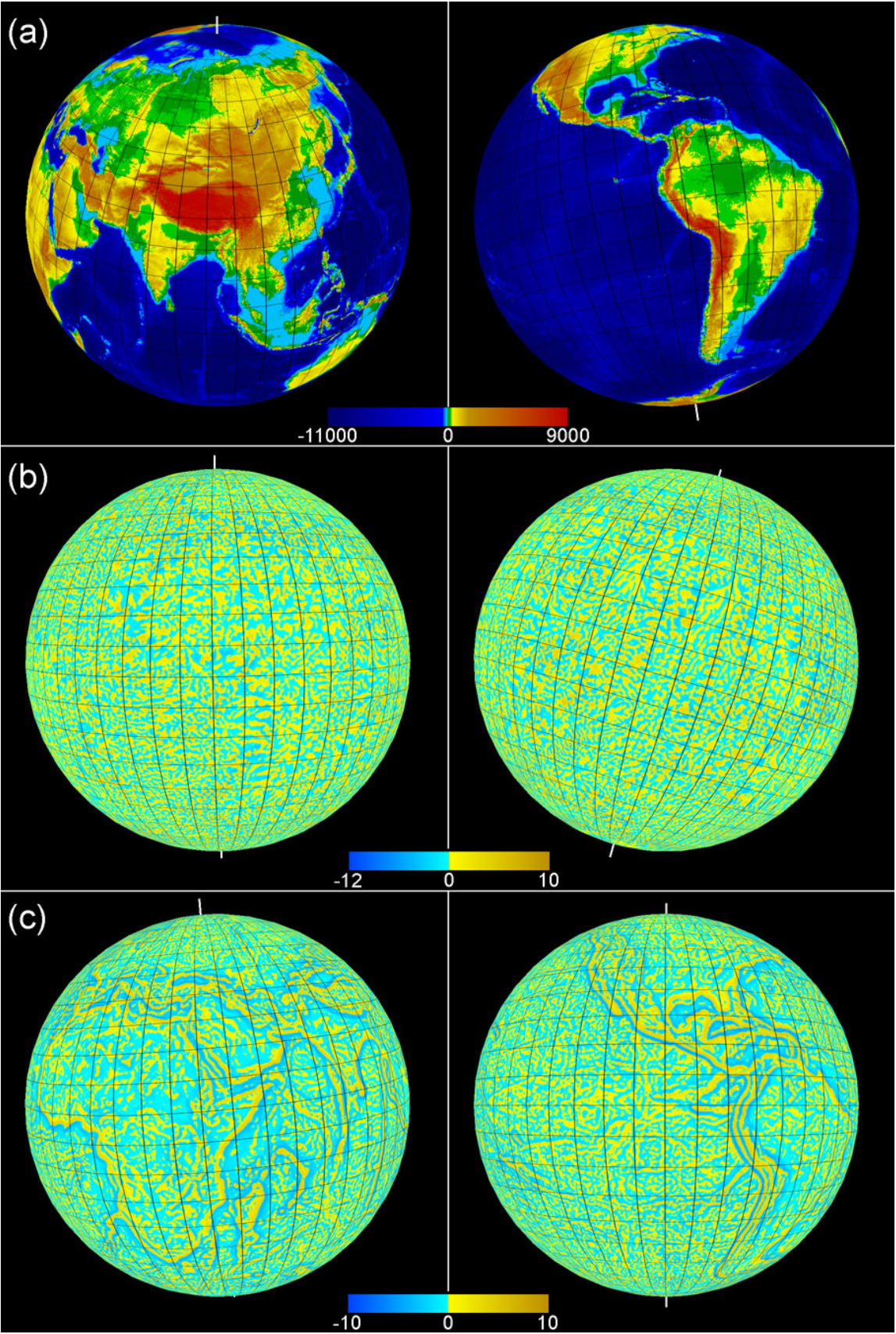

Fig. 5. The Earth morphometric globes: (a) elevation, (b) horizontal curvature, (c) vertical curvature, (d) minimal curvature, (e) maximal curvature, and (f) catchment area. Legends of all morphometric variables are in logarithmic scale except for elevation given in meters.

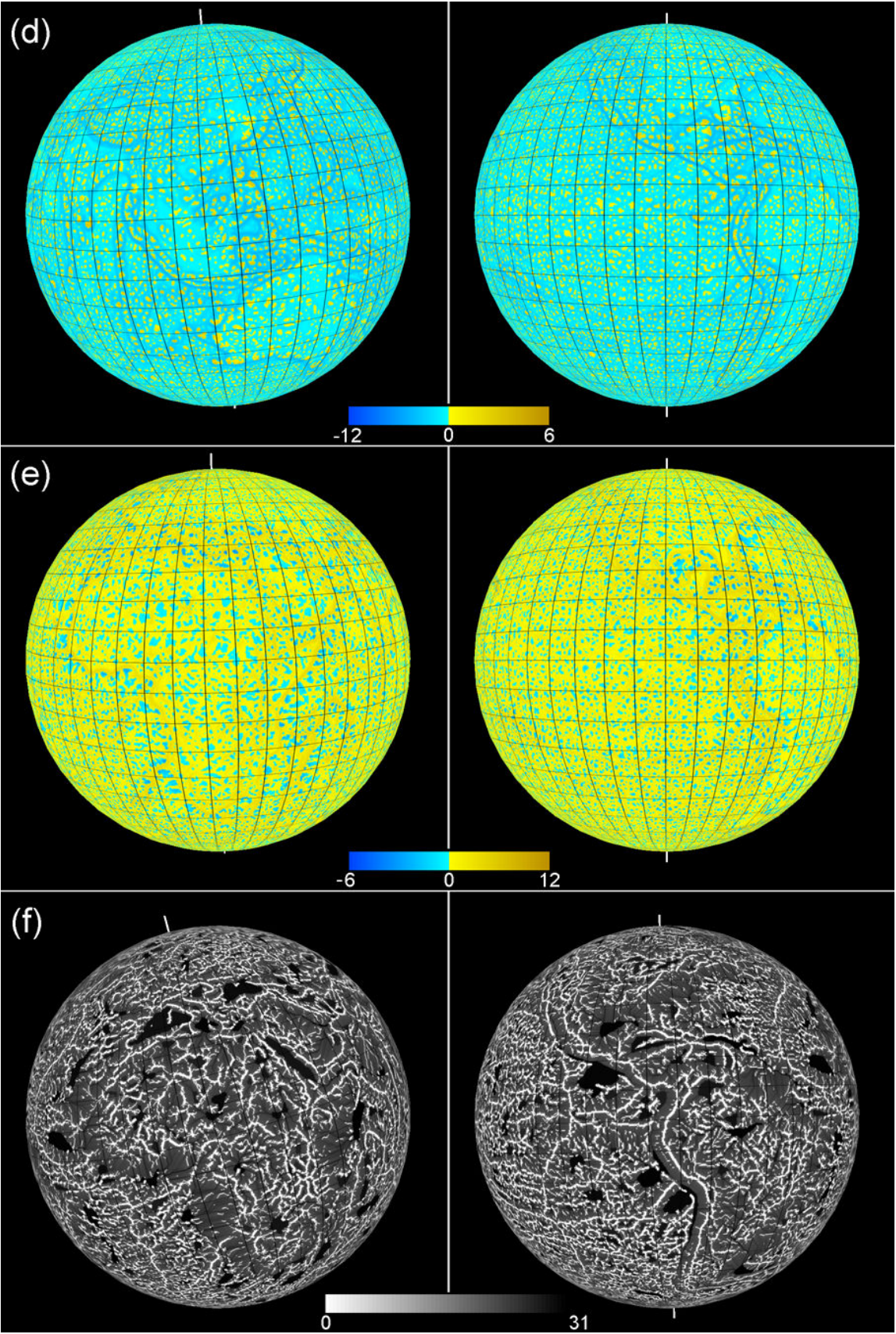

Fig. 5. (*continued*).

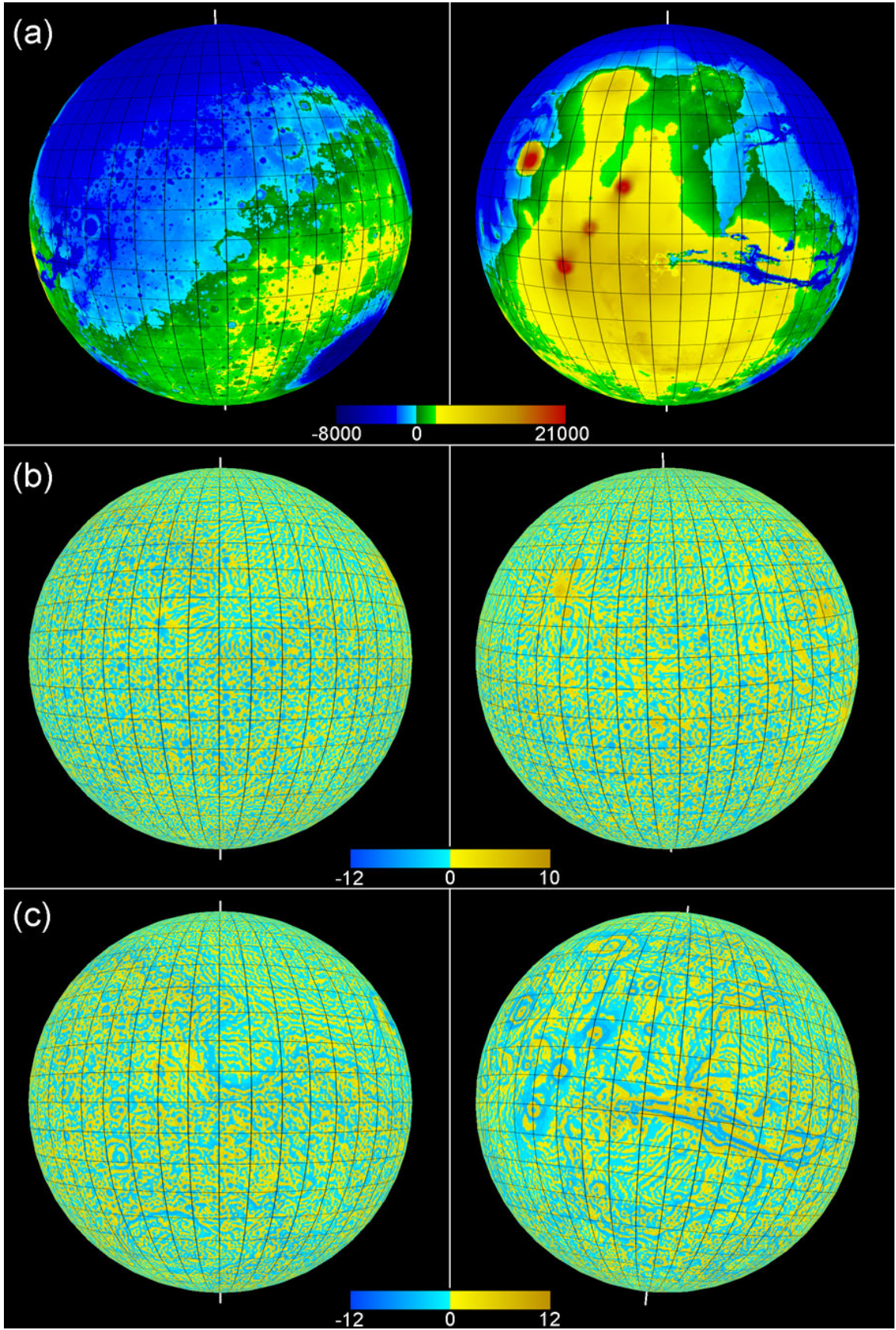

Fig. 6. The Mars morphometric globes: (a) elevation, (b) horizontal curvature, (c) vertical curvature, (d) minimal curvature, (e) maximal curvature, and (f) catchment area. Legends of all morphometric variables are in logarithmic scale except for elevation given in meter.

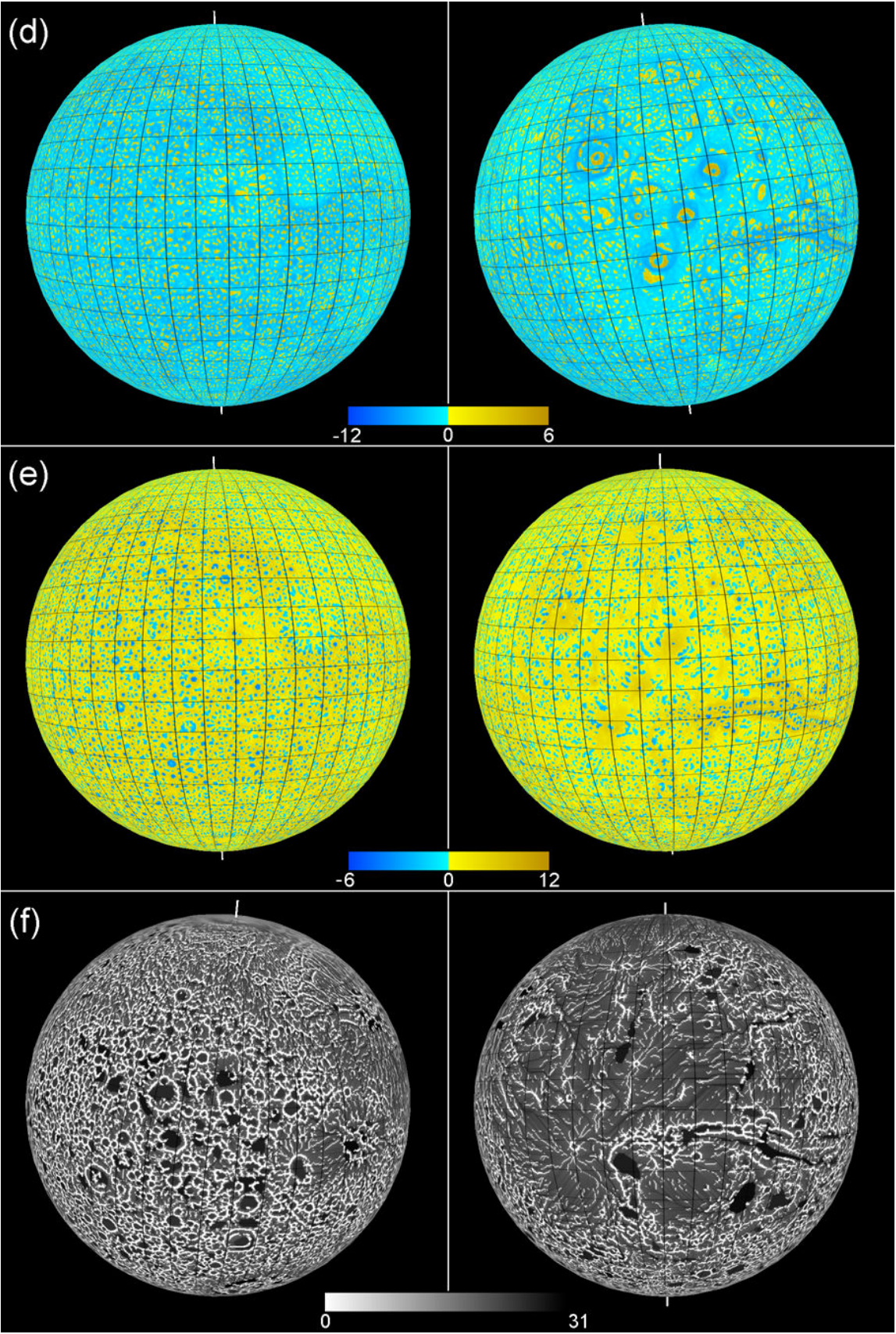

Fig. 6. (*continued*).

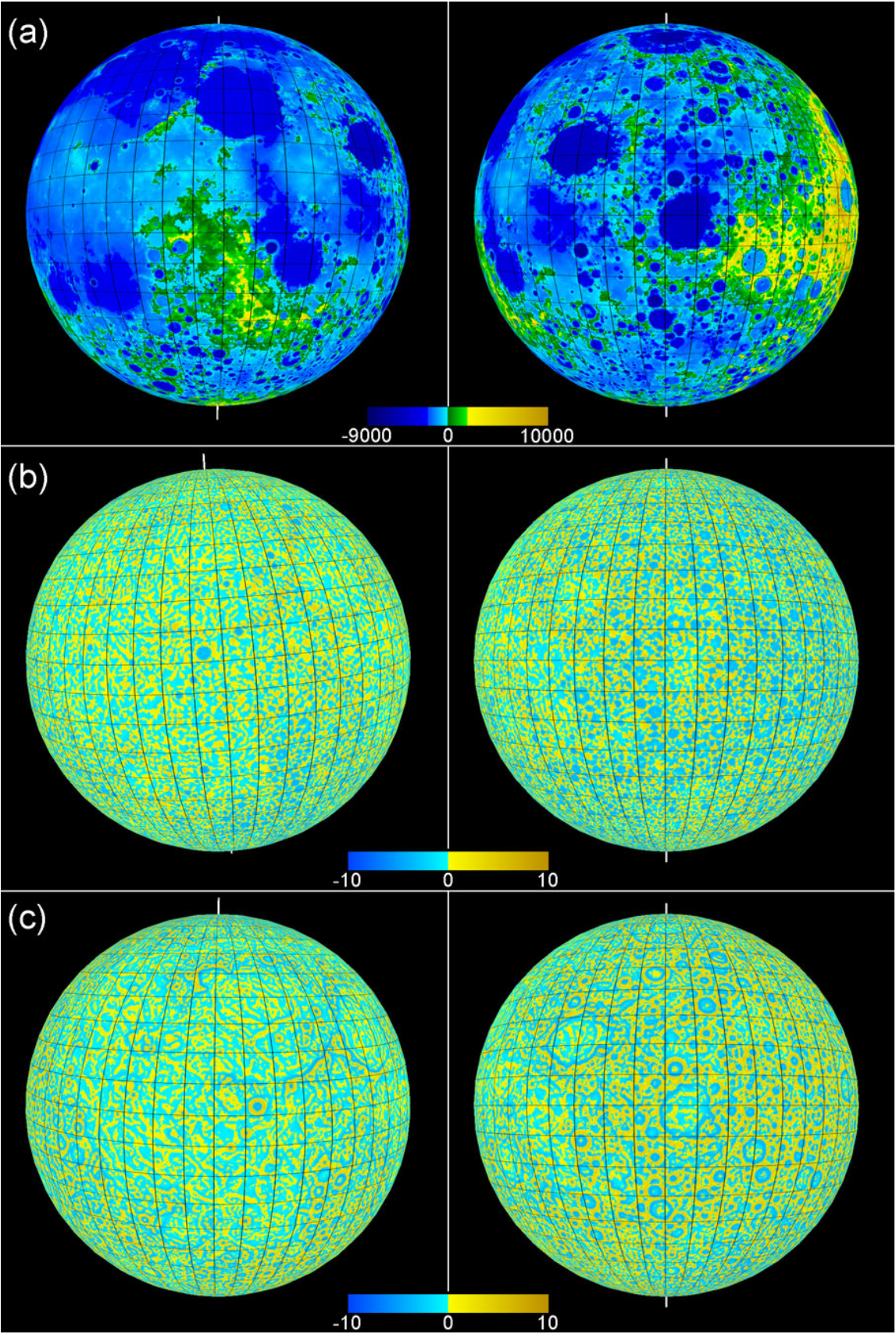

Fig. 7. The Moon morphometric globes: (a) elevation, (b) horizontal curvature, (c) vertical curvature, (d) minimal curvature, (e) maximal curvature, and (f) catchment area. Legends of all morphometric variables are in logarithmic scale except for elevation given in meters.

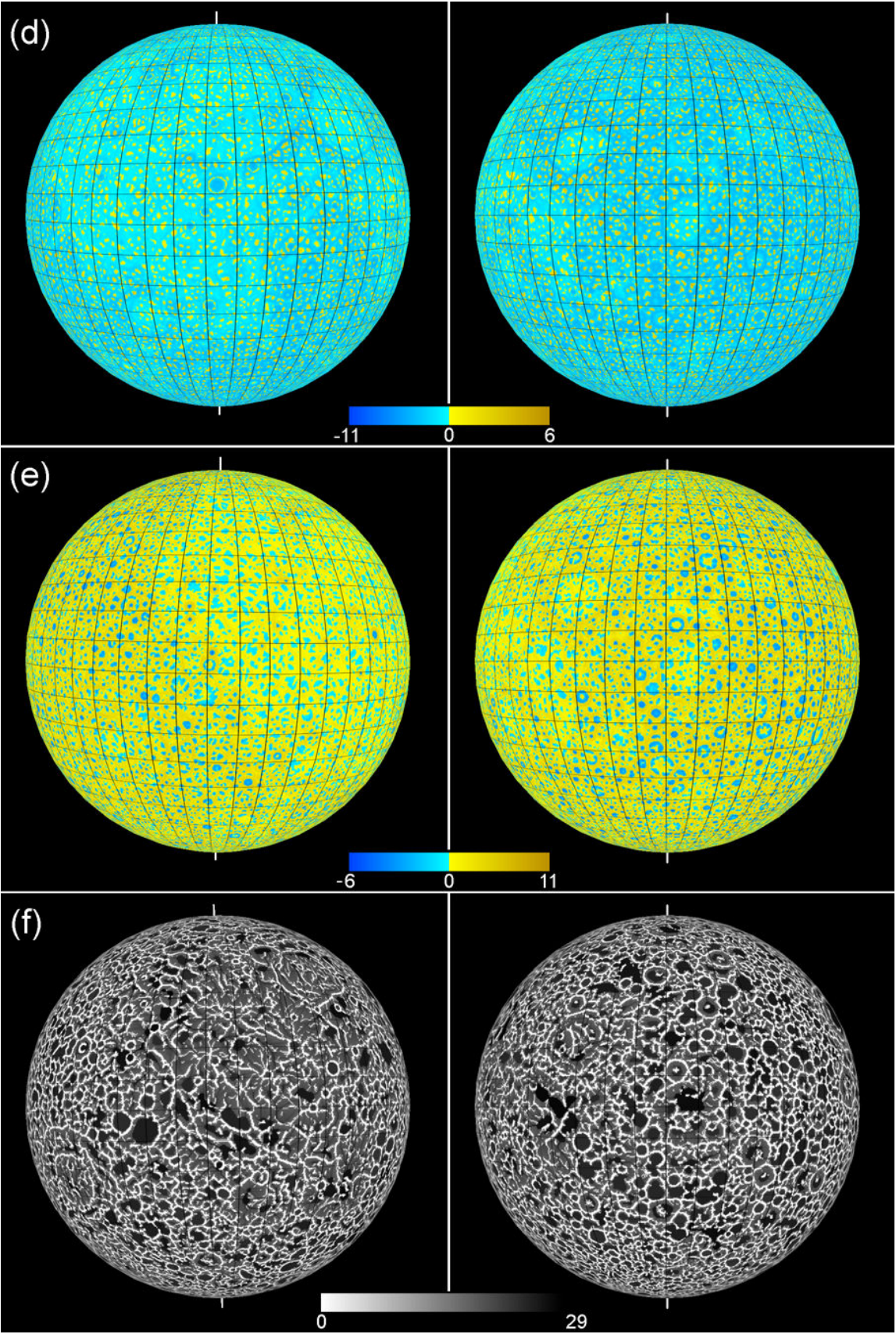

Fig. 7. (*continued*).

For the same angular resolution of 15', morphometric globes of the Earth, Mars, and the Moon have distinct linear resolution (around 27.6 km, 14.8 km, and 7.6 km on the equator, correspondingly) and scales. Morphometric globes clearly represent peculiarities of planetary topography, according to the physical and mathematical sense of a particular morphometric variable.

For example, horizontal curvature delineates areas of flow divergence and convergence (positive and negative values, correspondingly). These areas relate to spurs of valleys and ridges (blue and yellow patterns on the $k_h$ globes, respectively), which form so-called flow structures. For the Earth, they are most pronounced in ocean basins (Fig. 5b). On the $k_h$ globes of Mars (Fig. 6b), one can see a system of flow structures incoming to Utopia Planitia from Nilosyrtis and Protonilus Mensae and Elysium Planitia and Mons. For the Moon, the horizontal curvature represents cell-like patterns (Fig. 7b) resulting from a predominance of craters at the global scale.

Vertical curvature is a measure of relative acceleration and deceleration of flows (positive and negative values, correspondingly). Among other features, the $k_v$ globes of the Earth shows "mega-scarps", such as edges of continents and mountains (Fig. 5c). On the $k_v$ globes of Mars (Fig. 6c), one can see boundaries of Hellas Planitia, Isidis Planitia, Valles Marineris, foothills of Olympus Mons, Alba Patera, and so on. On the $k_v$ globes of the Moon (Fig. 7c), one can see well-marked boundaries of Mare Smythii, Mare Crisium, as well as a plethora of craters.

Catchment area measures an upslope area potentially drained through a point on the topographic surface. On the *CA* globes of the Earth (Fig. 5f), low values of *CA* delineate land and ocean ridges as white lines (e.g., the Andes and mid-ocean ridges). High values of *CA* show land valleys and ocean canyons as black lines, as well as land depressions and ocean basins as dark areas (e.g., the Mediterranean, Black, and Red Seas). On the *CA* globes of Mars (Fig. 6f), one can see the planetary network of valleys and canyons, as well as a large feature, Solis Planum, and a plethora of smaller depressions, predominantly craters.

Linear artifacts are typical for polar areas of all morphometric globes except for elevation. They were caused by both reduced accuracy of DEMs in polar areas and some peculiarities of global DEM processing.

## 4. Conclusions

We developed the first testing version of the system of virtual morphometric globes for the Earth, Mars, and the Moon. The system was constructed using Blender, the open-source software for 3D modeling and visualization. The real-time testing of the developed system demonstrated its good performance.

The first version of the system is as simple as possible: we used morphometric textures of low resolution (15'). Morphometric textures of higher resolution (up to 30") and hierarchical tessellation of the globe surface will be utilized in next versions of the system.

## Acknowledgements

The study was supported by RFBR grant 15-07-02484.